# Impact of Normal Lung Volume Choices on Radiation Pneumonitis Risk Prediction in Locally Advanced NSCLC Radiotherapy

Alyssa Gadsby, MS, Tian Liu, PhD, Robert Samstein, MD, Jiahan Zhang, PhD, Yang Lei, PhD, Kenneth E. Rosenzweig, MD, Ming Chao, PhD*

Department of Radiation Oncology, Icahn School of Medicine at Mount Sinai, New York, NY

*Corresponding Author:

Ming Chao, PhD

Email: ming.chao@mountsinai.org

Disclosure: None.

Funding: Radiation Oncology Institute (Grant #ROI 2022-9132)

## Acknowledgment

This manuscript was prepared using data from Datasets from the NCTN/NCORP Data Archive of the National Cancer Institute's (NCI's) National Clinical Trials Network (NCTN). Data were originally collected from clinical trial NCT number NCT00533949 "A Randomized Phase III Comparison of Standard-Dose (60 Gy) Versus High-Dose (74 Gy) Conformal Radiotherapy With Concurrent and Consolidation Carboplatin/Paclitaxel +/- Cetuximab (IND #103444) in Patients With Stage IIIA/IIIB Non-Small Cell Lung Cancer". All analyses and conclusions in this manuscript are the sole responsibility of the authors and do not necessarily reflect the opinions or views of the clinical trial investigators, the NCTN, or the NCI.

# Abstract

**Objectives:** To evaluate the impact of varying definitions of normal lung volume on the prediction of radiation pneumonitis (RP) risk in patients with locally advanced non-small-cell-lung-cancer undergoing radiotherapy.

**Materials and Methods:** Dosimetric variables V20, V5, and mean lung dose (MLD) were extracted from the treatment plans of 442 patients enrolled in the NRG Oncology RTOG 0617 trial. Three different definitions of lung volume were evaluated: total lung excluding gross-tumor-target, total lung excluding clinical-target-volume (TL-CTV), and total lung excluding planning-target-volume (TL-PTV). Patients were grouped as "no-RP2" (N = 377, grade $\leq$ 1 RP) and "RP2" (N = 65, grade $\geq$ 2 RP). Statistical analyses were performed to assess the effect on lung volume definition on RP2 prediction. Three supervised machine learning (ML) models—logistic regression (LR), k-Nearest Neighbor (kNN), and eXtreme Gradient Boosting (XGB)—were used to evaluate predictive performance. Model performance was quantified using the area under the receiver operating characteristic curve (AUC), and statistical significance was tested via a bootstrap analysis. Shapley Additive Explanations (SHAP) were applied to interpret feature contributions to model predictions.

**Results:** Statistical analyses showed that V20 and MLD were significantly associated with RP2, while differences among the lung volume definitions were not statistically significant. Both kNN and XGB classifiers consistently yielded higher AUC values for the TL-PTV definition compared to the other definitions, a finding supported by bootstrap analysis. SHAP analysis further indicated that V20 and MLD were the most influential predictors of RP2.

**Conclusion:** In line with previous studies, both statistical analysis and SHAP interpretation confirmed that V20 and MLD were associated with RP2. The machine learning models indicated that defining normal lung volume as total lung excluding PTV yielded the highest predictive performance for RP2 risk. Further validation using external datasets is warranted to confirm these findings.

# 1    Introduction

Treatment-induced lung injuries, such as radiation pneumonitis (RP), significantly limit the efficacy of radiotherapy (RT) [1]. Grade 2 or higher RP (RP2) is a major obstacle to the following consolidative immunotherapy, a standard treatment for patients with stage III non-small-cell-lung-cancer (NSCLC). Accurately predicting which patients will develop RP is crucial for timely prevention or intervention prior to onset of symptoms, and for planning subsequent therapies. Dosimetric parameters such as V20 (lung volume receiving > 20 Gy) and mean lung dose (MLD) are commonly used to assess and mitigate RP risk when evaluating RT treatment plans. Their values depend on the definition of 'normal lung volume' but no consensus exists across institutions.

In the era of three-dimensional conformal radiotherapy (3D-CRT), the normal lung volume was defined as the bilateral lung excluding the planning target volume (PTV) [2]. The RTOG 0617 trial later defined it as the bilateral lung volume excluding the clinical target volume (CTV) [3]. More recent clinical trials, such as RTOG 0618, 0813, 0915, 1106 and ESTRO-ACROP guidelines recommended using the lung volume excluding gross tumor volume (GTV) [4]. Compared to 3D-CRT, intensity modulated radiation therapy (IMRT) technique has become the preferred technique in treating NSCLC. Due to its non-uniform dose distribution outside the target, significant variations in dosimetric parameters are expected across different lung volume definitions.

Several studies have examined the impact of different normal lung volume definition on RP prediction. One study involving 100 patients with stage I-III NSCLC assessed four lung volume definitions and found significant differences in MLD and V20 across these definitions. Notably, MLD calculated with the GTV exclusion showed improved accuracy in predicting RP [5]. Another study, with 183 lung cancer patients treated with IMRT, compared RP prediction performance using V5 (lung volume receiving > 5Gy), V20, and MLD derived from three lung volume definitions [6]. Results revealed considerable differences in dosimetric parameters, suggesting that excluding PTV from total lung volume may enhance RP prediction accuracy. Most recently, a study [7] analyzing 117 patients with stage III NSCLC treated with chemotherapy concurrent or sequential volumetric modulated arc therapy (VMAT) indicated that MLD using the GTV exclusion method was more predictive of RP2, aligning with the findings of previous studies [5].

The key takeaway from these studies is the ongoing lack of consensus on the optimal lung volume definition for predicting RP risk. Additionally, many of these studies apparently were limited by insufficient data, which hindered the ability to draw definitive conclusions. Furthermore, the datasets used were imbalanced, yet no measures were implemented, potentially leading to statistical inaccuracies. Lastly, previous analyses relied solely on logistic regression models, which may not fully capture the complex relationships with the dataset. Consequently, we believe it is important to conduct further research using comprehensive methods and larger datasets to address the limitations of previous studies. In this work, we analyzed data from the publicly available RTOG 0617 trial, which included over 500 patients with locally advanced NSCLC. We employed three supervised machine learning (ML) classifiers for RP2 prediction and applied data balancing techniques to mitigate class imbalance issues. To gain deeper insights into feature contributions, we utilized SHAP (SHapley Additive exPlanations) [8] to interpret the model outputs.

## 2   Materials & Methods

### 2.1   Patient Data Preparation

A subset of 442 patients were selected for this study from 544 patients enrolled in RTOG 0617 [3, 9] by excluding those with incomplete data. All patient data including tabulated data in several spreadsheet files, data dictionaries, planning computed tomography (CT), RT structure, and RT dose were downloaded from The Cancer Imaging Archive (TCIA) [10] with permission. RP grades, particularly acute grades, were used in this analysis. A binary classification was adopted for the clinical endpoints: "no-RP2" (N = 377) for patients with acute RP grades 0 or 1, and "RP2" (N = 65) for patients with acute grades 2 or higher. The planning CT images, RT structures, and RT doses for all patients were exported into an open source software platform CERR [11] and were converted into MATLAB® (MathWorks, Natick, MA) format files for data mining. Three lung volumes were generated using CERR if they were not already present. These volumes were defined as follows: the total lung (Total Lung) excluding the gross tumor volume, TL excluding the clinical target volume (Total Lung-CTV), and TL excluding the planning target volume (Total Lung-PTV). Three sets of dosimetric variables—MLD, V5, V20—were calculated based on the three lung volume definitions and exported, along with the binary endpoint data, for subsequent statistical analysis and ML prediction studies.

### 2.2   Statistical Analysis

Three statistical methods were used to examine the correlations between RP2, lung volumes, and dosimetric variables: repeated analysis of variance test (ANOVA), the Mann-Whitney U test, and univariate logistic regression. To prepare the data and check the statistical capabilities, Shapiro-Wilk test was applied to verify the statistical suitability of V5, V20, and MLD distributions that were extracted for each of the three lung volumes. The test confirmed suitability by examining if the variable is normally distributed. ANOVA provides insights into differences (statistical variance) between groups of data. The Mann-Whitney U test determined differences in V5, V20, and MLD variables with a dependence on non-RP2 and RP2 definitions. Univariate logistic regression was used to find connections between RP2 and each of V5, V20, and MLD separately.

## 2.3 Machine Learning Prediction and SHAP Explanation

The effect of the normal lung volume definitions on RP2 prediction was further investigated with three supervised ML classifiers taking MLD, V5, and V20 as inputs: logistic regression (LR), k-Nearest Neighbor (kNN), and eXtreme Gradient Boosting (XGB) using the scikit-learn library with Python v3.8.10. The entire dataset was split according to the 80:20 scheme, where 80% of data was reserved for model training and validation while 20% was for testing. Prior to model training and validation, these models underwent hyper-parameter tuning with five-fold cross-validation using grid search based on the training dataset [12]. The performance for each ML classifier was assessed with the area under the curve (AUC) of the receiver operating characteristic (ROC) from ten-fold cross validation using the training data. To further evaluate prediction accuracy and AUC, a bootstrap analysis with 1,000 iterations was conducted.

The original dataset was seriously imbalanced, which could compromise the ML model performance. To mitigate class imbalance, we utilized a technique called synthetic minority over-sampling technique (SMOTE) [13] to increase the minority (RP2) group to match the majority (non-RP2) group in the training data but the test data remained the same. Comparison was made between the results obtained with the resampled balanced dataset and those from the original imbalanced dataset.

To understand the influence of each dosimetric variable on ML-based RP2 prediction, we employed the explainable SHAP technique [8]. SHAP values assign a numerical value to each variable or feature in a model, showing how much that feature affects the model's prediction for a given data point. They can also show the significance of each feature compared to others, and, how much the model relies on interactions between features.

## 3 Results

Shapiro-Wilk tests showed that the V5 variable, across all three lung volume definitions, had p-values greater than 0.05, indicating it followed a normal distribution. In contrast, both V20 and MLD exhibited a quasi-normal distribution ($p < 0.05$). The distributions of these variables, categorized into two classes (RP2 vs non-RP2), are shown in Fig. S1 of the Supplementary Material. Although the t-test would be more appropriate for V5, we used the Mann-Whitney U test to compare differences between the two classes for all three variables. The interquartile ranges (IQRs) in the box-and-whisker plots (Fig. S2) demonstrate almost complete overlap between the two classes across all different volume definitions. Figure 1 presents the results of three statistical methods (Table S1 in the Supplementary Material lists the numerical values of these methods). ANOVA tests, based on p-values, revealed significant differences in V20 and MLD between the two lung volumes, while V5 showed no significant difference (Fig. 1a). Mann-Whitney U tests exhibited a clear statistical difference between RP2 and non-RP2 groups for V20 and MLD, while no significant difference was observed for V5 ($p > 0.05$, Fig 1b). Univariate logistic regression (Fig. 1c) indicated a descending predictive strength among the dosimetric variables—V20, MLD, and V5—based on decreasing odds ratios. V20 showed the strongest association with RP2 risk, while V5 was the least predictive ($p > 0.05$). This analysis also suggests that the association with RP2 is independent of lung volume definitions, as the odds ratios for each dosimetric variable remain comparable across different definitions.

Figure 2 shows the ROC curves generated from ten-fold cross validation using three supervised classifiers, based on the original imbalanced dataset (top row) and the resampled balanced dataset (bottom row). The corresponding AUC values are listed in Table 1. All three classifiers demonstrated comparable performance on both the imbalance and balanced datasets. However, among them, the LR classifier yielded the lowest AUC values and showed no notable differences across the three lung volume definitions. XGB consistently showed differences among three lung volumes in both datasets, with the AUC for the Total Lung-PTV volume substantially higher than those for the other two volume definitions. A similar pattern was observed in kNN, suggesting that the Total Lung-PTV definition may be more suitable for RP2 prediction in the context of the RTOG 0617 trial data.

The results of bootstrap analysis for prediction accuracy and AUC performance are presented in Fig. 3. The LR classifier showed consistently low prediction accuracy across the three lung volume definitions, whereas both the kNN and XGB classifiers achieved higher and comparable accuracy. A similar pattern was observed for AUC values, although kNN outperformed XGB in this metric. Detailed statistics including the mean, standard deviation, 50th percentile, 75th percentile, and maximum values are listed in Table S2 of the Supplementary Material.

Figure 4 illustrates the mean SHAP values obtained using the test data for each classifier across three lung volume definitions. Positive SHAP values (shown in red) indicate a positive correlation with pneumonitis risk, whereas negative values (in blue) indicate a negative correlation. For LR (Figs. 4a-4c), the mean SHAP values across all three lung volumes are close to zero, suggesting that none of the variables significantly contributed to RP2 prediction, consistent with the previously reported ROC/AUC results. In contrast, kNN (Figs. 4d-4f) shows that all three variables contributed to RP2 prediction with similar magnitude. For XGB (Figs. 4g-4i), V20 emerged as the strongest predictor of RP2, followed by MLD (Fig. 4g), while V5 contributed the least.

# 4    Discussion

Evaluating the impact of different lung volume definitions on pneumonitis risk prediction is crucial for determining whether variations in volume selection influence the predictive power of dosimetric variables on RP risk. This research aims to address this clinical question. Unlike previous studies, our investigation utilized a dataset four times larger and applies more comprehensive analytical approaches. While our statistical analysis findings are consistent with earlier studies, the machine learning classifiers trained on data from the RTOG 0617 trial suggest that the lung volume definition excluding PTV may be more predictive of RP risk than the other two definitions. This observation is supported by results from both the kNN and XGB classifiers. As shown in Fig. 2 and Table 1, the AUC values of this lung volume definition are statistically higher than those of the other two definitions across both imbalanced (Figs. 2b and 2c) and balanced datasets (Figs. 2e and 2f). Additional support for this finding comes from the bootstrap analysis shown in Fig. 3. These results are consistent with the volume definition used in 3D-CRT studies [2] and further confirm the findings reported in [6]. However, it is important to note that these results were derived from a single source data and relied on three dosimetric variables. Further testing and external validation are necessary before these findings can be integrated into clinical practice.

Both the Mann-Whitney U test and univariate logistic regression analysis indicate that V20 and MLD are statistically significant for classifying RP2 risk, while V5 is not. This result partially aligns with previous findings [5, 6] but contradicts the findings reported in [14, 15]. For example, [15] reported that lung V5 ≥ 40% was associated with RP2, whereas [14] concluded that none of these variables were predictive of RP2. In contrast, our ML classifiers, particularly kNN and XGB, suggested that these variables may be effective in predicting RP2. Although these classifiers do not inherently reveal variables importance, SHAP analysis (Fig. 4) provided insight into their relative contributions. Specifically, V20 and MLD emerged as potentially effective predictors, despite their relatively small mean SHAP values. Additionally, the SHAP analysis using the kNN classifier indicated that MLD and V5 may also contribute to RP2 prediction. Discrepancies between our findings and earlier studies may stem from various factors, including differences in treatment planning techniques (e.g. 3D-CRT vs. IMRT/VMAT), the use of concurrent chemotherapy, dataset size, and analytical approaches. Notably, our analysis did not stratify patients based on treatment modality or chemotherapy status, which may have introduced variability in model training.

Moreover, the reliance on V20, V5, and MLD alone may not represent the optimal approach for RP2 risk modeling. As suggested in [16-18], incorporating spatial information of dose variations within the lung could enhance the accuracy of toxicity prediction. Several studies have successfully integrated such spatial features into their predictive modeling for toxicities in lung and head and neck cancers [18-23]. For instance, [20] reported that in patients with radiation-induced lung damage the dose to the peripheral medial-basal portion of the lungs was consistently higher. While incorporating this level of detail is beyond the scope of current study, it represents an important direction for future research. Nevertheless, our findings highlight underscore the importance of robust, large-scale, and stratified studies to resolve these inconsistencies and establish consensus on the most predictive dosimetric variables and lung volume definitions for clinical use.

To address whether class imbalance contributed to reduced ML performance, we applied SMOTE to generate a balanced training dataset by oversampling the minority class. The test dataset remained unchanged to ensure unbiased evaluation. With this balanced dataset, both kNN and XGB classifiers demonstrated improved performance (Figs 2b-2c vs. 2e-2f), whereas LR classifier showed no noticeable enhancement. Several factors may explain this discrepancy. First, LR assumes a linear relationship between input variables and the RP2 outcome, which may not be true in this context. In contrast, kNN and XGB are capable of modeling complex, non-linear relationships, allowing them to better capture the underlying data patterns. Second, despite the mitigation of class imbalance, LR may still be sensitive to the quality of representation within the minority class, particularly if oversampling introduces noise or fail to sufficiently capture variability. Third, LR does not inherently model interactions between variables unless explicitly defined, whereas models like XGB capture such interactions naturally through their decision tree structures, an advantage especially evident in more balanced datasets. Finally, LR typically has higher bias and limited flexibility, which could be more pronounced following oversampling. On the other hand, XGB has lower bias and is more effective at managing the tradeoff between bias and variance, leading to better performance.

The rationale for using SHAP to interpret model outputs is to identify which variables contribute most significantly to RP2 prediction, especially given the ongoing lack of consensus regarding the relative importance of V20, V5, and MLD. As shown in Fig. 4, although the mean SHAP values are relatively small, they still provide insights into the sensitivity of each variable in relation to RP2 prediction across different classifiers. Consistent with the ROC/AUC analysis, the LR classifier showed no meaningful predictive power for any of these variables, suggesting a poor model fit. In contrast, V20 consistently emerged as the most influential feature across different lung volume definitions, while the contributions of MLD and V5 varied. For the kNN classifier (Figs. 4d-4f), all three variables contributed to RP2 prediction with comparable magnitudes. In the case of the XGB classifier (Figs. 4g-4i), V20 was the dominant predictor, followed by MLD while V5 had minimal impact. Although SHAP supports additional visualization tools, such as summary plots and dependence plots, which can aid in interpreting feature contributions and model behavior, we did not include them here due the overall small SHAP values and limited predictive strength of the variables, which reduce the potential for meaningful or informative interpretation from such plots.

# 5   Conclusion

Our analysis, consistent with most previous studies, indicates that V20 and MLD are associated with RP2, as supported by both the statistical methods and SHAP analysis. Machine learning classifiers further suggest that the lung volume definition excluding the PTV may be more predictive of RP2 risk than the other two definitions, as demonstrated by the ROC/AUC results. However, further investigation using external datasets is necessary to validate these findings and to confirm the most appropriate lung volume definition and predictive variables for clinical application.

# Caption of Figures

**Figure 1**. Statistical analyses of three dosimetric variables V5, V20, and MLD calculated from three lung volume definitions. (a) Analysis of Variance (ANOVA) test; (b) Mann-Whitney U test; (c) Univariate logistic regression. The p-values are colored in red for values less than 0.05 and in black for values greater than 0.05, for ease of illustration. Abbreviation: TL – Total Lung; CTV – clinical target volume; PTV – planning target volume.

**Figure 2**. Receiver operating characteristic (ROC) curves from 10-fold cross validations with three supervised machine learning classifiers: Logistic Regression, k-Nearest Neighbor, and eXtreme Gradient Boosting. Top row: using the original imbalanced dataset; Bottom row: using the resampled balanced dataset. Abbreviation: CTV – clinical target volume; PTV – planning target volume.

**Figure 3**. Box-and-whisker plots from the bootstrap analysis (1,000 iterations) showing prediction accuracy (a) and areas under the receiver operating characteristic curves (AUC) (b) for the three classifiers across the three lung volume definitions. Abbreviation: CTV – clinical target volume; PTV – planning target volume.

**Figure 4**. Mean SHAP values for each of the dosimetric variables that influences the RP prediction with three classifiers: Logistic Regression (top row), k-Nearest Neighbor (middle row), and eXtreme Gradient Boosting (bottom row) using the test data.

# Caption of Table

**Table 1**. Areas under the receiver operating characteristic curves from ten-fold cross-validation with three classifiers based on the imbalanced (rows of no shading) and balanced datasets (rows of gray shading), respectively. Abbreviation: CTV – clinical target volume; PTV – planning target volume.

# Figure 1.

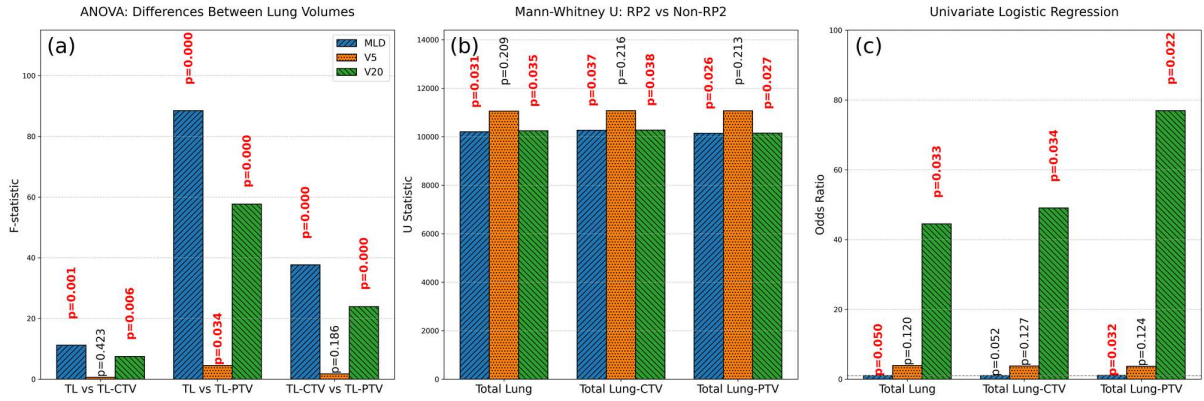

# Figure 2.

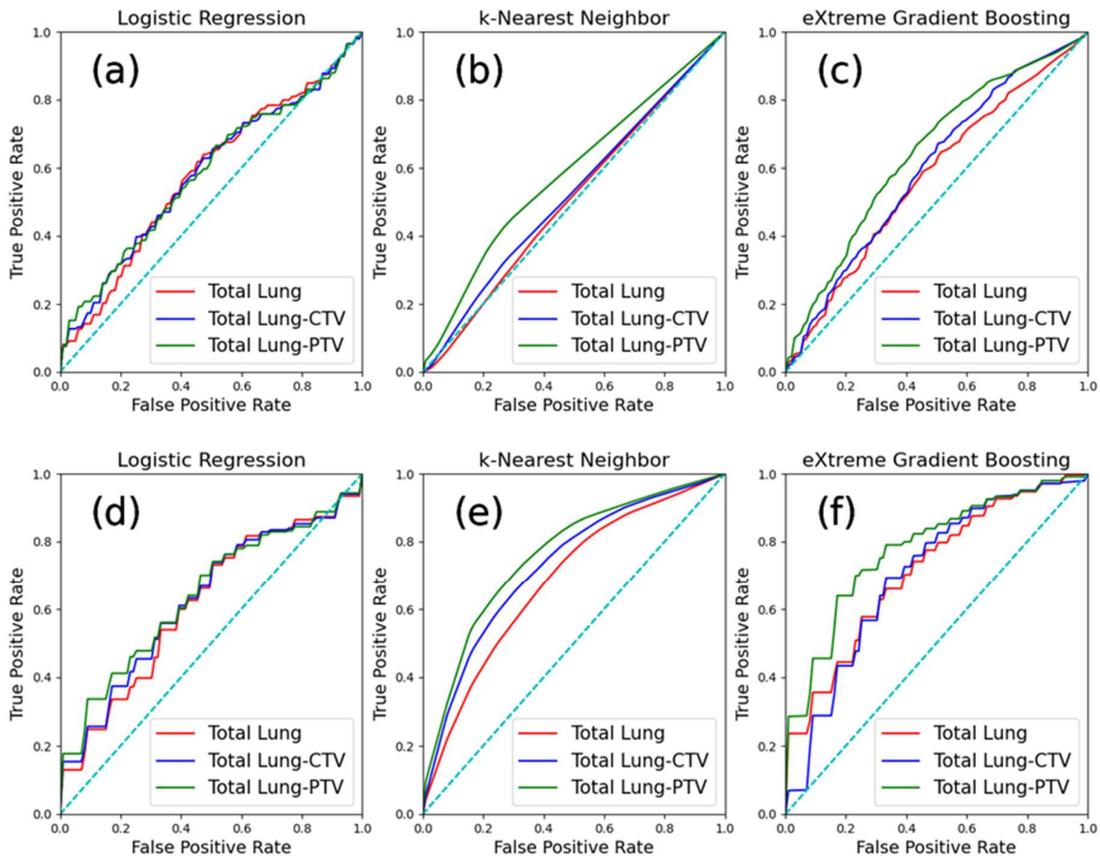

**Figure 3.**

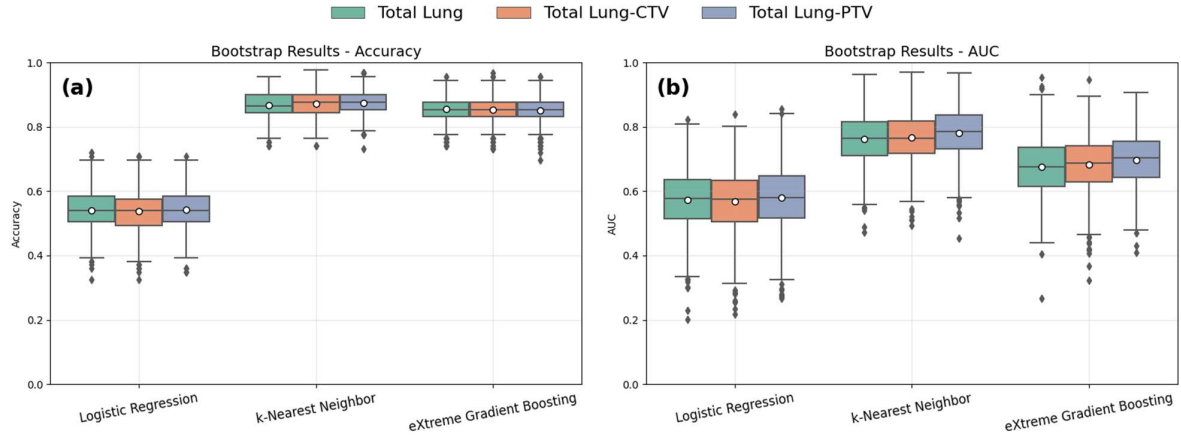

**Figure 4.**

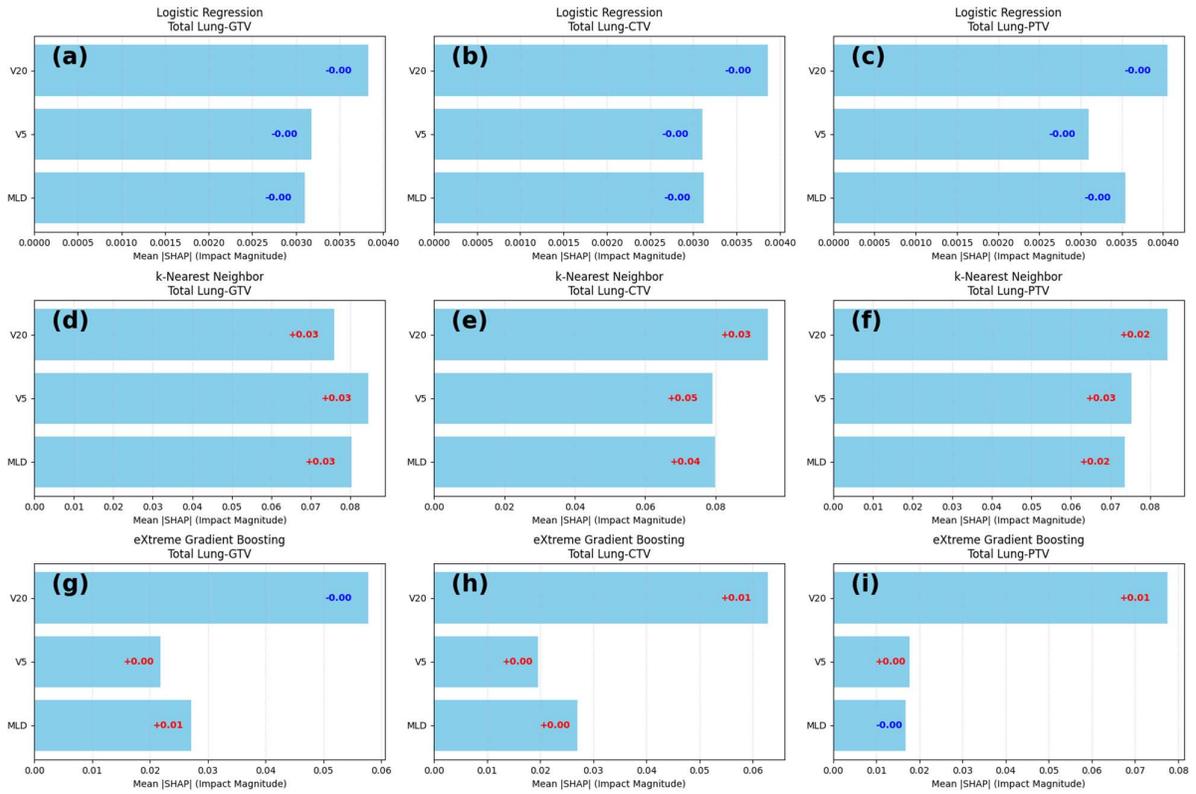

**Table 1.**

|  |  | Total Lung | Total Lung - CTV | Total Lung - PTV |
|---|---|---|---|---|
| Logistic Regression Classifier (LR) | Imbalanced Data | 0.579 ± 0.127 | 0.582 ± 0.131 | 0.583 ± 0.131 |
|  | Balanced Data | 0.616 ± 0.124 | 0.628 ± 0.123 | 0.643 ± 0.114 |
| k-Nearest Neighbor Classifier (kNN) | Imbalanced Data | 0.508 ± 0.082 | 0.525 ± 0.109 | 0.584 ± 0.108 |
|  | Balanced Data | 0.685 ± 0.102 | 0.729 ± 0.099 | 0.760 ± 0.089 |
| eXtreme Gradient Boosting (XGB) | Imbalanced Data | 0.573 ± 0.128 | 0.598 ± 0.117 | 0.644 ± 0.093 |
|  | Balanced Data | 0.715 ± 0.088 | 0.704 ± 0.092 | 0.780 ± 0.129 |